\documentclass[
     reprint,
     superscriptaddress,
     amsmath,amssymb,amsfonts
     aps,
     pra,
     floatfix,
    ]{revtex4-1}

\usepackage{booktabs}
\usepackage{xfrac}
\usepackage{braket}
\usepackage{graphicx} \graphicspath{{PIC}}
\usepackage{verbatim}
\usepackage{etoolbox}
\usepackage[utf8]{inputenc}
\usepackage{mathtools}
\usepackage{hyperref} \hypersetup{colorlinks=true,citecolor=blue,linkcolor=blue,urlcolor=blue}
\usepackage{csquotes}
\usepackage{multirow}
\usepackage{longtable}
\usepackage{resizegather}
\usepackage[normalem]{ulem}
\usepackage{orcidlink}
\usepackage{placeins}

\usepackage{bm}
\usepackage{array}[=2016-10-06]
\newcolumntype{C}[1]{>{\centering\arraybackslash}p{#1}}
\newcolumntype{Y}[1]{>{\centering\arraybackslash}m{#1}}

\usepackage{color}
\usepackage{morefloats}

\definecolor{Gray}{gray}{0.9}
\definecolor{LightCyan}{rgb}{0.88,1,1}
\definecolor{green}{rgb}{0.5451,0.2706,0.0745}

\def\DFT{{\small  DFT}}

\def\VASP{{\small VASP}}

\begin{document}

\title{Mixing and Preserving Properties of Non-van der Waals 2D Materials in Sandwich-like Heterostructures}

\author{Anastasiia Nihei\,\orcidlink{0009-0006-0851-709X}}
\affiliation{Theoretical Chemistry, Technische Universit\"at Dresden, 01062 Dresden, Germany}
\affiliation{Institute of Ion Beam Physics and Materials Research, Helmholtz-Zentrum Dresden-Rossendorf, 01328 Dresden, Germany}
\author{Rico Friedrich\,\orcidlink{0000-0002-4066-3840}}
\email[]{rico.friedrich@tu-dresden.de}
\affiliation{Theoretical Chemistry, Technische Universit\"at Dresden, 01062 Dresden, Germany}
\affiliation{Institute of Ion Beam Physics and Materials Research, Helmholtz-Zentrum Dresden-Rossendorf, 01328 Dresden, Germany}

\date{\today}

\begin{abstract}
\noindent
Heterostructure interfaces created by stacking two-dimensional (2D) materials outline an appealing route towards achieving advanced electronic and magnetic functionalities at the nanoscale.
Non-van der Waals (non-vdW) sandwich-like heterostructures (HSs) are build out of non-vdW 2D materials derived from non-layered crystals and establish a new frontier enabling strong interfacial coupling.
Employing \emph{ab initio} density functional theory (DFT), we investigate two classes of such trilayer HSs: purely non-vdW assemblies and hybrid vdW/non-vdW/vdW stacks.
We show that non-vdW sandwiches undergo pronounced geometric reconstruction forming chemically bonded interfaces with large absolute binding energies up to an order of magnitude larger than those of conventional vdW HSs.
These strong interactions drive extensive electronic hybridization, band gap renormalization, and tunable interlayer magnetic coupling.
In contrast, vdW encapsulation largely protects the intrinsic atomic and electronic structure of the non-vdW sheet from the environment
, while enabling charge transfer doping of the vdW layers.
Our results thus establish sandwich HSs a versatile interface design platform, spanning interaction regimes from weakly perturbed encapsulation to strongly interacting chemically bonded systems with emergent electronic and magnetic behavior.

\vspace{0.2cm}

\end{abstract}

\maketitle

\section{Introduction} \label{intro}

The rapid development and processing of two-dimensional (2D) materials has enabled the fabrication of van der Waals (vdW) heterostructures (HSs), where atomically thin layers are vertically stacked and held together by weak dispersion forces \cite{Geim_Nature_2013,Novoselov_Science_2016}. These HSs provide a versatile platform for engineering electronic band alignment
\cite{Liu_NatRevMat_2016} and interlayer charge-transfer processes, including the formation and control of interlayer excitons \cite{Rivera_NatNano_2018}.
They also give access to moiré physics upon twisting \cite{Cao_Nature_2018,Arnold_MatTod_2024} while mostly preserving the distinct characteristics of the constituent layers.
However, the weak interlayer coupling inherent to vdW systems fundamentally limits the degree of electronic and magnetic hybridization, making strongly coupled interfacial phenomena such as substantial charge transfer, enhanced magnetic coupling, and emergent correlated states more difficult to realize \cite{Gibertini_NNano_2019}.

In contrast, non–van der Waals (non-vdW) 2D materials \cite{Balan_MatTod_2022,Kaur_AdvMat_2022}, including ultrathin oxides as well as other binary and ternary compounds derived from strongly bonded non-layered bulk crystals, offer a fundamentally different approach to interface engineering.
Unlike conventional vdW materials, non-vdW compounds do not possess chemically saturated weakly vdW bonded but more reactive cleavage planes~\cite{Barnowsky_NanoLett_2024}.
This makes their isolation into atomically thin sheets more challenging, but at the same time promotes them as appealing precursors to achieve strong interfacial HS coupling.

Recent experimental advances have indeed made it possible to manufacture non-vdW 2D materials already for a range of several dozen mostly ionic (oxidic) systems
\cite{Guan_AdvMat_2017,Tai_advmat_2017,Li_acsnano_2018,Puthirath_Balan_NNANO_2018,Puthirath_Balan_CoM_2018,Yadav_AdvMatInt_2018,Puthirath_Balan_acsanm_2018,Wang_JPCL_2020,Zhang_angchem_2019,Xie_ACSAMI_2019,Liu_JMCA_2019,Feng_matter_2020,Puthirath_Small_2020,Guo_Nanoscale_2021,Ouyang_NML_2021,Hu_SmallStruct_2021,Goktuna_PSCE_2021,Peng_NChem_2021,Toksumakov_NPJ2DM_2022,Jiang_NatSyn_2023,Mahapatra_2DM_2023,Chattopadhyay_NL_2024,Su_advsci_2024,
Kaur_ACSNano_2020,Puthirath_JPCC_2021,Moinuddin_AdvElMat_2021,Hu_two-dimensional_2021,Yousaf_JPCC_2021,Gibaja_AdvMat_2021,Xu_ChemSci_2021,Homkar_ACSApplNanoMat_2021,Chahal_ACSML_2021,Zhang_NMAT_2021,Chen_FlatChem_2022,Puthirath_Balan_OxOpMatSci_2023}
including 2D WO$_3$, hematene (2D Fe$_2$O$_3$), and ilmenene (2D FeTiO$_3$).
Additional computational works even suggest a potential materials space of several thousand systems~\cite{Jia_npjcm_2021,Friedrich_NanoLett_2022,Barnowsky_AdvElMats_2023,Hu_NatComm_2023,Bagheri_JPCL_2023,Bjork_Science_2024,Zhong_npjcm_2025,Ono_PRB_2025,Bagheri_CM_2026,Barnowsky_NatCommun_2026}.
Following an initial focus on liquid phase exfoliation,
subsequent studies have established a broader range of approaches for producing non-vdW 2D materials, including chemical exfoliation, molecular beam epitaxy, chemical vapour deposition, and other thin-film growth techniques \cite{Balan_MatTod_2022,Jiang_NatSyn_2023,Ren_2023_CVD_MBE}.
This demonstrates that the absence of conventional vdW cleavage planes does not preclude the realization of atomically thin materials, but rather motivates alternative manufacturing strategies.

The increasing experimental accessibility of non-vdW 2D materials as well as current rising interest in interfaces of stacked (oxidic) thin films~\cite{Li_AdvMat_2022,Li_AdvMat_2022,Shen_ACSAMI_2022,Sanchez_Nature_2024,Sha_NatComm_2024,Wang_AdvMat_2024,Schmidbauer_AplPhysLett_2025} motivates their integration into HSs.
When two non-vdW materials are brought into contact, chemical bonding and charge redistribution at the interface can produce hybridized electronic states and strong interlayer magnetic exchange, giving rise to properties that are inaccessible in weakly coupled vdW HSs \cite{Kou_JPCL_2017,Nihei_ActaMat_2026}.

\begin{figure*}[ht!]
    \includegraphics[width=0.9\textwidth]{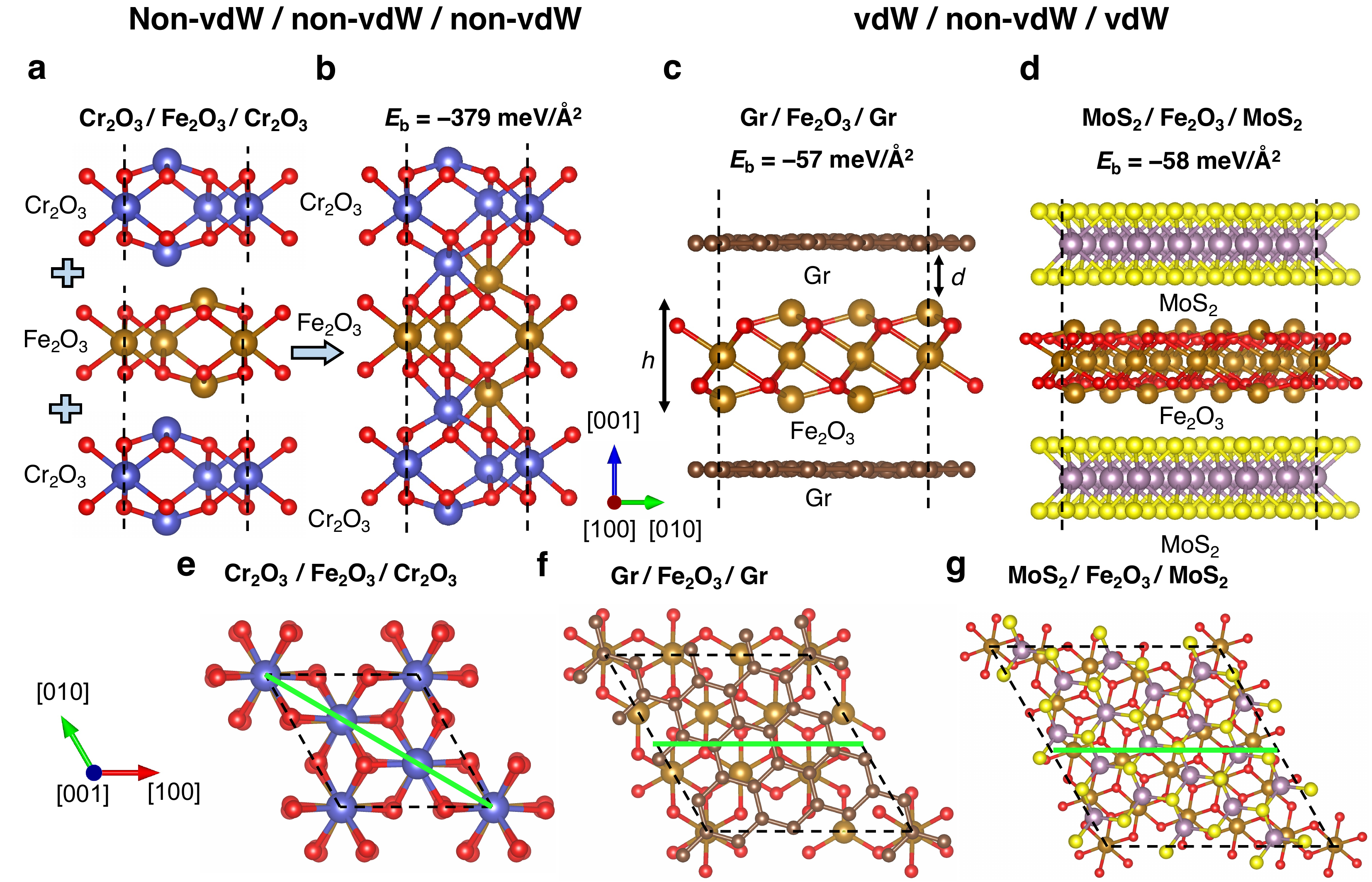}
    \caption{\small \textbf{Interfacial bonding in different types of sandwich-like HSs.} (\textbf{a}) Visual representation of initial non-vdW 2D sheets used for constructing the (\textbf{b}) purely non-vdW sandwich structures (such as Cr$_2$O$_3$/Fe$_2$O$_3$/Cr$_2$O$_3$). (\textbf{c,d}) Optimized structure of hybrid vdW/non-vdW/vdW sandwiches (Gr/Fe$_2$O$_3$/Gr, MoS$_2$/Fe$_2$O$_3$/MoS$_2$). Binding energies ($E_{\mathrm{b}}$) per unit area are shown for all three systems, confirming the substantially stronger bonding in non-vdW sandwiches compared to vdW/non-vdW/vdW hybrids. The inner layer thickness $h$ and interlayer distance $d$ are shown for the example of Gr/Fe$_2$O$_3$/Gr in (\textbf{c}). Gr stands for graphene. The compass in the middle of the upper panels indicating crystal directions refers to subfigures (\textbf{a-d}).
    (\textbf{e-g}) Top views of the representative sandwich-like HSs. The position of the planes on which the charge density differences are visualized is indicated by the green lines. The compass in the bottom left of the panels indicating crystal directions refers to subfigures (\textbf{e-g}).
    Atom colors: Cr, blue; Fe, gold; O, red; C, brown; Mo, violet; S, yellow.}
    \label{fig1}
\end{figure*}

Trilayer, \emph{i.e.}, sandwich-like HSs represent a particularly promising architecture for exploring interfacial coupling.
Compared to bilayers, trilayers introduce two materials interfaces and confine the central layer, providing additional pathways for enhanced interlayer electronic and magnetic interactions.
Hereafter, the terms “layer” and “trilayer” are used for simplicity to also describe the constituent sheets and three-sheeted sandwich structures of non-vdW HSs, respectively, although a discrete layer is not necessarily well defined in non-vdW materials.

In this work, we perform a systematic first-principles investigation of sandwich-like HSs constructed from non-vdW oxides and other 2D compounds.
We compare purely non-vdW trilayers with hybrid vdW/non-vdW/vdW stacks, focusing on structural relaxation, interfacial bonding, charge redistribution, electronic hybridization, and interfacial magnetic exchange coupling.
The non-vdW sandwich HSs show strong interfacial bonding as already evidenced by the optimized structures and quantified by the bonding energies.
These are a factor three to eight larger than the ones of hybrid structures placing the non-vdW compound between vdW layers.
This finding is further supported by charge density difference analysis where only for the former systems appreciable charge reorganization over the whole interface regions is identified.
The different degree of interaction is also visible in the computed band structures where substantial hybridization is found for the strongly bonded non-vdW sandwiches.
This finally translates also to a sizable magnetic coupling across the interfaces.
However, when placing a diamagnetic sheet in the middle of the sandwich structure, the outer layers can be effectively decoupled enabling switching between configurations in external magnetic fields.

\section{Results and Discussion} \label{results_discussion}

\textbf{Construction and Structural Relaxation.} Sandwich-like (trilayer) HS selection is guided by the results obtained for the heterobilayer systems of non-vdW 2D materials~\cite{Nihei_ActaMat_2026}.
We focus primarily on binary compounds such as Fe$_2$O$_3$ and Cr$_2$O$_3$, whose corresponding bilayer HSs exhibit pronounced emerging interfacial properties such as hybrid interface bands and strong magnetic coupling.
The sandwich HSs are constructed as detailed in the Methods section, using two complementary design strategies to investigate distinct interfacial interaction regimes. In the first approach, purely non-vdW sandwiches are formed by stacking three non-vdW layers, with identical materials forming the outer layers (e.g. Cr$_2$O$_3$/Fe$_2$O$_3$/Cr$_2$O$_3$, as presented in Fig.\ref{fig1}(\textbf{a})).
In the second approach, hybrid vdW/non-vdW/vdW HSs are created by encapsulating a non-vdW monolayer between two vdW sheets, specifically graphene or MoS$_2$.
For MoS$_2$/Fe$_2$O$_3$/MoS$_2$, HSs with different twist angles (4$^{\circ}$, 17$^{\circ}$, 42$^{\circ}$, and 55$^{\circ}$) are investigated.
All systems were initially assembled with a layer separation ($d$) of 2 \AA\ and subsequently relaxed using the SCAN-rVV10 functional \cite{Perdew_SCAN_PRL_2015,Sabatini_rVV10_PRB_2013,Peng_PRX_2016} allowing for a unified description of both covalent/ionic bonding and dispersive interactions.
All relaxed structures obtained using the two approaches are presented in Fig.~S1 (purely non-vdW HSs) and Fig.~S2 (hybrid vdW/non-vdW/vdW HSs) in the SI.

Upon structural relaxation, the two classes of HSs exhibit fundamentally different behavior. In purely non-vdW sandwiches, the initially separated layers undergo
pronounced reconstruction, converging toward a bulk-like geometry with the formation of interfacial covalent and/or ionic bonds (see Fig.~\ref{fig1}(\textbf{b})).
We note that the maximum strain acquired by one sheet per direction in the relaxed structure is 2.55\% for Rh$_2$O$_3$ in Cr$_2$O$_3$/Rh$_2$O$_3$/Cr$_2$O$_3$.
The relaxed atomic configurations deviate strongly from those of the corresponding freestanding monolayers and initial structures, with substantial changes in coordination environments, bond lengths, and interlayer spacing. Specifically, cation–oxygen bond lengths at the interfaces contract to $2.1-2.3$ \AA , while the local coordination of interfacial transition-metal ions changes from threefold to sixfold.
In addition, an increase in the inner layer thickness ($h$, indicated next to Fig.~\ref{fig1}(\textbf{c})) of up to 0.6~\AA\ with respect to the bilayer HS \cite{Nihei_ActaMat_2026} and up to 1~\AA\ with respect to individual 2D sheets are observed \cite{Friedrich_NanoLett_2022}, indicating partial bulk-like reorganization of the atoms.
These structural distortions suggest that stacking non-vdW sheets drives the system toward a new thermodynamic minimum resembling a strongly coupled multilayer assembly, in which interfacial bonding plays a dominant role in stabilizing the HS.

In contrast, hybrid vdW/non-vdW/vdW sandwiches show minimal structural distortion upon relaxation (Fig.~\ref{fig1}(\textbf{c,d})).
The non-vdW sheet largely retains its freestanding geometry, and also the encapsulating vdW layers preserve their characteristic planar structures.
The interlayer distances $d$ shown in Table \ref{tab:summary}  are, however, smaller ($2.48-2.84$ \AA) in comparison to those typical for vdW homobilayers and HSs ($>3$ \AA ) \cite{Pakdel_NCom_2024, Cao_Nature_2018,Arnold_MatTod_2024}, suggesting that the interaction is dominated by dispersion forces with some minor contributions from electrostatic or weak orbital overlap effects. This structural robustness highlights encapsulation as a potential viable strategy for stabilizing non-vdW 2D materials without fundamentally altering their intrinsic atomic structure and properties.

\begin{figure}[ht!]
    \includegraphics[width=0.5\textwidth]{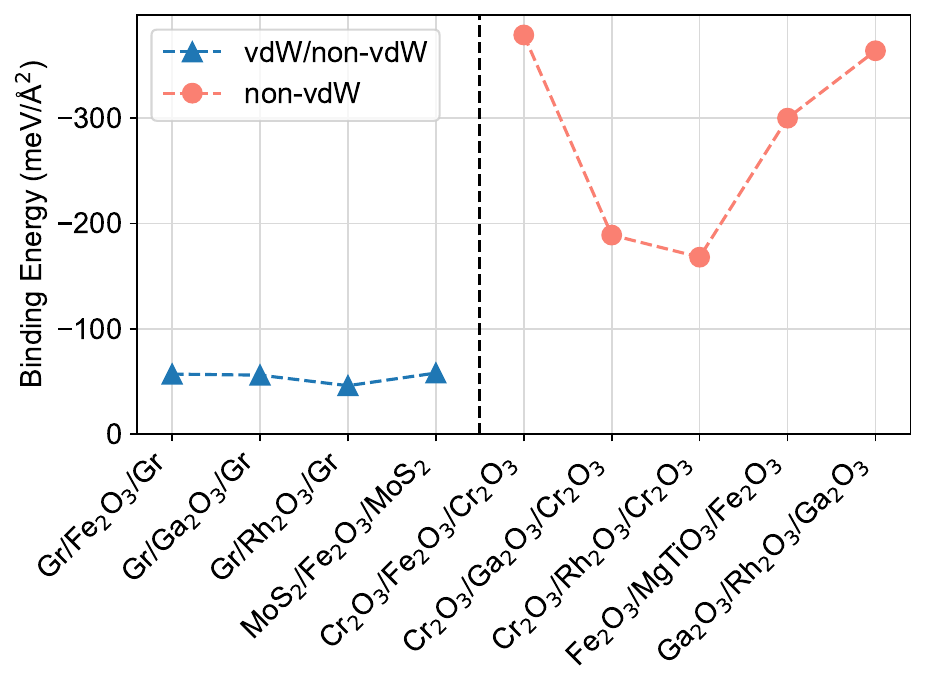}
    \caption{\small \textbf{Binding energies.} The left side shows binding energies of hybrid vdW/non-vdW/vdW HSs, whereas the right side refers to purely non-vdW HSs. The two classes are separated by a vertical dashed line.
    }
    \label{fig2}
\end{figure}

\textbf{Interlayer Bonding: From Weak to Strong Interactions.}
These observations are quantitatively supported by the calculated binding energies (Fig.~\ref{fig2}).
The binding energy, $E_{\mathrm{b}}$, is according to equation~\ref{Eb} in the Methods section defined as the difference between the total energy of the relaxed HS and the sum of the energies of the corresponding isolated, fully relaxed constituent 2D components, normalised per area.
By construction, $E_{\mathrm{b}}$ is negative, reflecting the energy gain upon HS bonding.
Purely non-vdW sandwiches exhibit strong interfacial coupling, with absolute binding energies ranging from 168 to 380~meV/\AA$^2$ per cell corresponding to 84-190~meV/\AA$^2$ per interface.
They are thus up to almost an order of magnitude larger than pure vdW interactions of $\sim$20~meV/\AA$^2$~\cite{Zacharia_PRB_2004,Mounet_AiiDA2D_NNano_2018,Jung_NanoLett_2018}.
Binding energies for the whole set of calculated sandwich-like HSs as well as interlayer bonding type and interlayer distance are shown in Table S1 in the SI.
The values of the non-vdW stacks are characteristic of chemically bonded interfaces and are comparable in magnitude to reported absolute binding energies of strongly coupled non-vdW heterobilayers (up to 232~meV/\AA$^2$ for Cr$_2$O$_3$/Rh$_2$O$_3$ \cite{Nihei_ActaMat_2026} recomputed on the SCAN+rVV10 level) as well as to exfoliation energies of non-vdW 2D materials \cite{Friedrich_NanoLett_2022,Barnowsky_AdvElMats_2023}.

\begin{table}[h!]
\centering
\caption{Summary of key structural and energetic parameters for selected sandwich-like HSs including bonding type, inter-layer distances, and binding energy.
    The two values for Cr$_2$O$_3$/Fe$_2$O$_3$ correspond to the interlayer distance of each of the interfacial cations to the O anions of the neighboring sheet. \vspace{0.1cm}}
\label{tab:summary}
\resizebox{0.5\textwidth}{!}{
\begin{tabular}{|>{\centering\arraybackslash}m{2.0cm}|>{\centering\arraybackslash}m{2.1cm}|>{\centering\arraybackslash}m{2.0cm}|>{\centering\arraybackslash}m{2.0cm}|}
\toprule
\hline
\textbf{Outer/ Inner layer} & \textbf{Interlayer Bonding Type} & \textbf{Interlayer Distance $d$ (\AA)} & \textbf{Binding Energy $E_{\mathrm{b}}$ (meV/\AA$^2$)} \\
\midrule
\hline
\vspace{0.1cm}
Cr$_2$O$_3$/Fe$_2$O$_3$ & Covalent/Ionic & 1.44, 1.47 & $-379$ \\
Gr/Fe$_2$O$_3$                  & van der Waals  & 2.57 & $-57$ \\
MoS$_2$/Fe$_2$O$_3$        & van der Waals  & 2.47 & $-58$ \\
\bottomrule
\hline
\end{tabular}
}
\end{table}

In comparison, hybrid vdW/non-vdW/vdW sandwiches display significantly weaker interfacial bonding, with absolute binding energies of approximately 46-59~meV/\AA$^2$ per cell corresponding to 23–30~meV/\AA$^2$ per interface.
These values slightly exceed typical vdW HS binding energies ($\sim$20~meV/\AA$^2$ \cite{Nihei_ActaMat_2026}), placing these systems at a slightly elevated interaction regime.
This moderate binding strength is consistent with the minimal structural distortion observed in these HSs and reflects predominantly dispersive interactions supplemented by weak electrostatic/covalent contributions.

\begin{figure}[ht!]
    \includegraphics[width=0.5\textwidth]{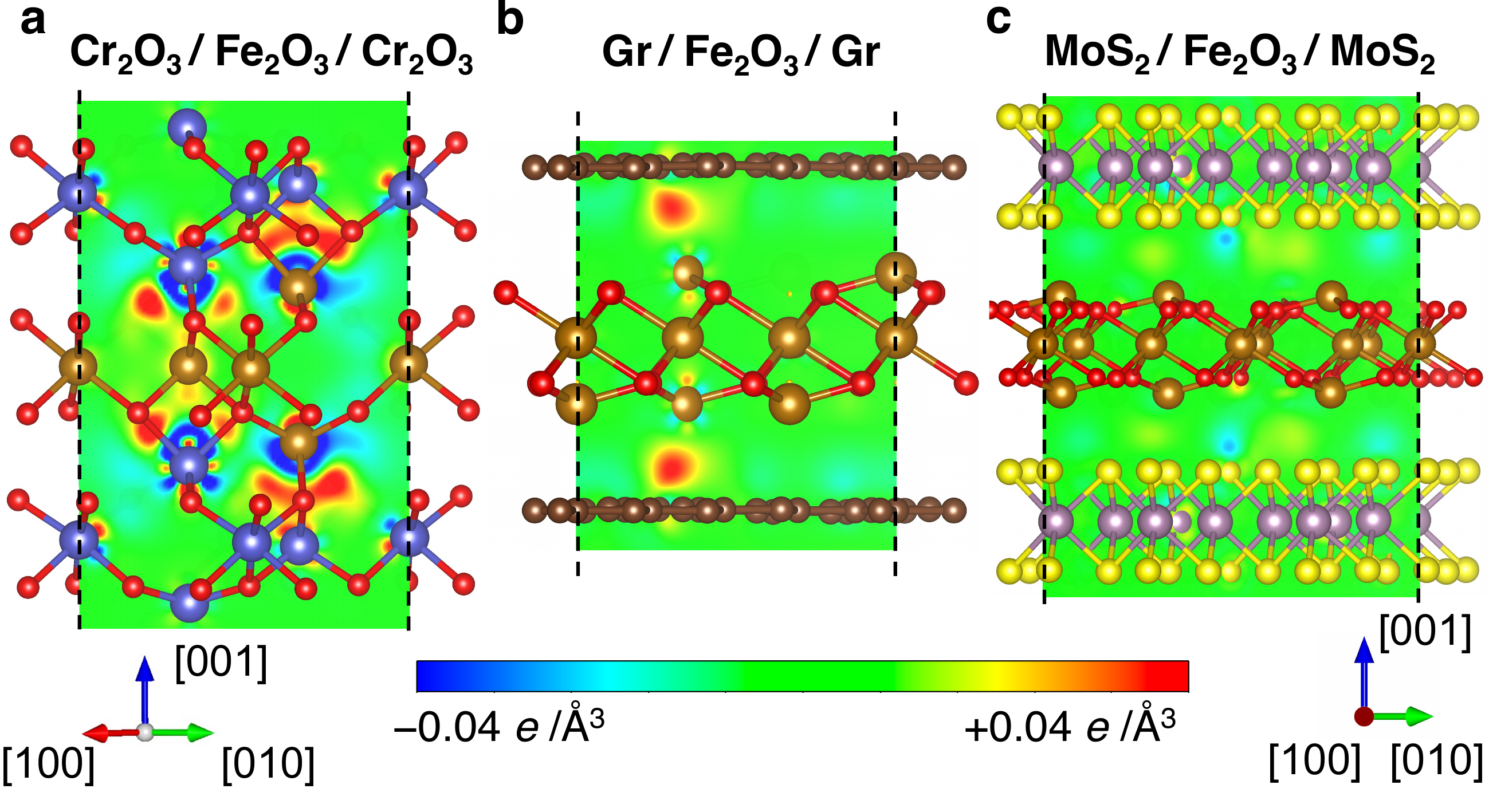}
    \caption{\small \textbf{Charge density differences.}  Charge density differences for (\textbf{a}) Cr$_2$O$_3$/Fe$_2$O$_3$/Cr$_2$O$_3$, (\textbf{b}) Gr/Fe$_2$O$_3$/Gr and (\textbf{c}) MoS$_2$/Fe$_2$O$_3$/MoS$_2$.
    The position of each plane on which the data is visualized is shown in the top views in Fig. \ref{fig1}(\textbf{e-g}).
     The color scale quantifies charge gain (red) and charge loss (blue).
     Dashed lines indicate the periodic boundaries of the unit cells.
     The compass on the left refers to (\textbf{a}), while the compass on the right refers to (\textbf{b}) and (\textbf{c}).}
    \label{fig3}
\end{figure}

Charge-density-difference analysis provides further microscopic insight into the nature of the interfacial bonding.
In purely non-vdW sandwiches (Fig.~\ref{fig3}(\textbf{a})), pronounced charge redistribution is observed around the interfaces -- characterized by alternating regions of charge accumulation around oxygen anions and depletion around cations extending across the interfacial bonds.
These findings are indicative of strong chemical interactions and substantial orbital hybridization, consistent with the formation of covalent and/or ionic bonds.
This behavior
confirms the large absolute binding energies and is in line with the extensive reconstruction observed upon relaxation.

Conversely, hybrid vdW/non-vdW/vdW sandwiches exhibit substantially weaker charge redistribution.
As shown in Fig.~\ref{fig3}(\textbf{b}), the charge-density difference of Gr/Fe$_2$O$_3$/Gr is characterized by modest electron accumulation on the vdW layers, accompanied by a corresponding spread out depletion in the adjacent non-vdW layer.
This redistribution corresponds to a small net charge transfer to graphene across the interface, driven by interfacial electrostatics.

For other  hybrid vdW/non-vdW/vdW systems exemplified by MoS$_2$/Fe$_2$O$_3$/MoS$_2$ in Fig.~\ref{fig3}(\textbf{c}), the charge rearrangement is minor and does not lead to pronounced charge accumulation in the interlayer region or the formation of directional charge bridges between atoms.
The internal charge distribution and bonding topology of the non-vdW layer remain intact, indicating the absence of orbital hybridization.
These features are consistent with an interaction regime dominated by dispersion forces, supplemented by weak electrostatic polarization effects.
In general, vdW encapsulation thus protects the non-vdW sheet from environmental influences
, despite the potential presence of some minor interfacial charge transfer.

\textbf{Electronic Structure: Mixing versus Preservation of Properties}.
The nature of interlayer bonding determines the electronic properties of the HSs.
Compared to individual monolayers and bilayer HSs, sandwich-like HSs generally exhibit enhanced interlayer electronic coupling due to the presence of two materials interfaces and the confinement of the central layer.
As a result, electronic hybridization effects can be amplified in trilayer assemblies, although the extent of hybridization strongly depends on the chemistry and bonding strength at the interfaces.

\begin{figure}[ht!]
    \includegraphics[width=0.5\textwidth]{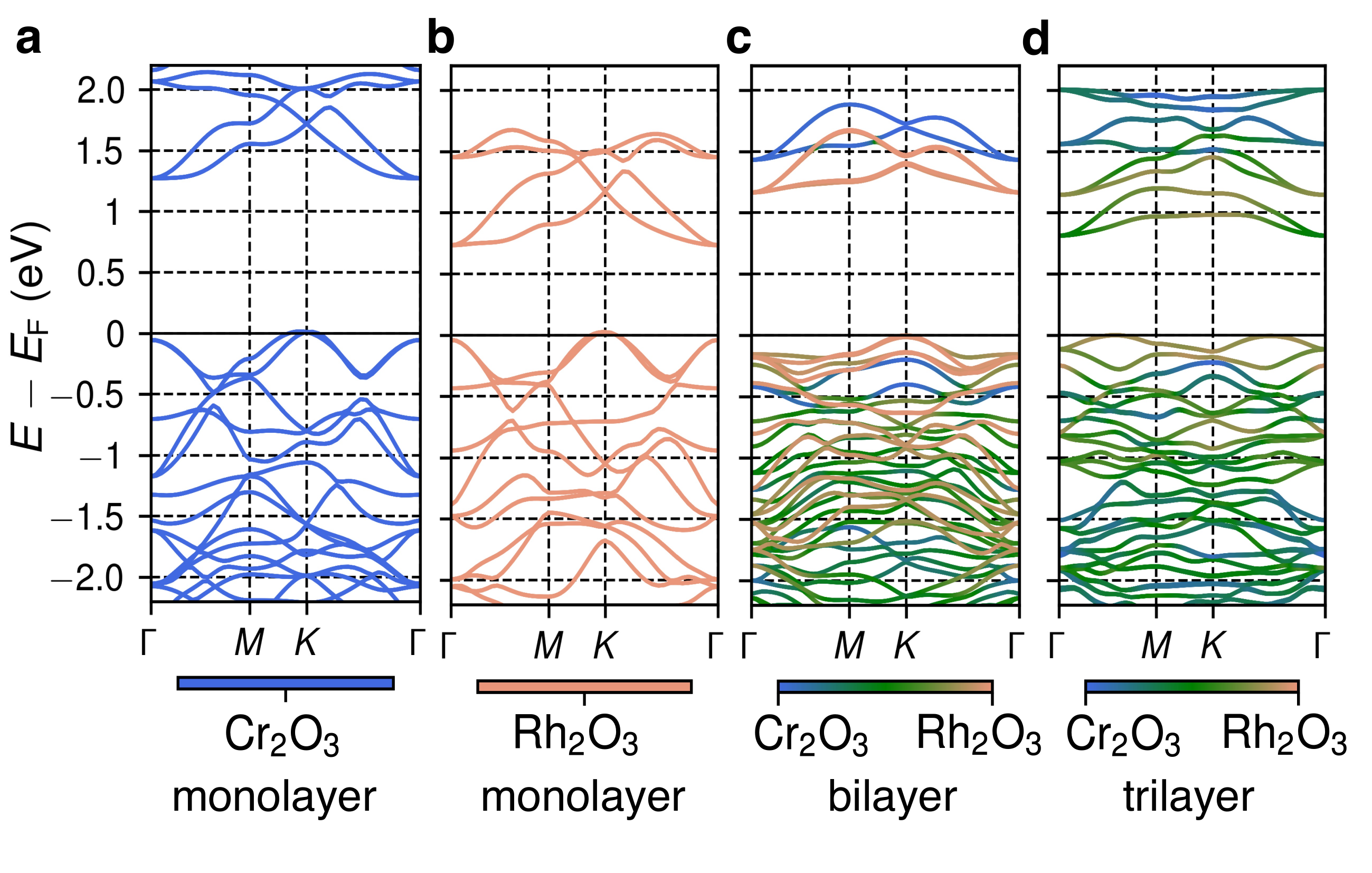}
    \caption{\small \textbf{Comparison of the projected band structures.} (\textbf{a,b}) Band structure of individual Cr$_2$O$_3$ and Rh$_2$O$_3$ monolayers.
    (\textbf{c}) Bilayer Cr$_2$O$_3$/Rh$_2$O$_3$ band structure with projection contributions onto the atoms of each subsystem indicating hybrid states by intermediate (green) color.
    (\textbf{d}) Trilayer Cr$_2$O$_3$/Rh$_2$O$_3$/Cr$_2$O$_3$ band structure.
    }
    \label{fig4}
\end{figure}

In purely non-vdW sandwiches with interfacial coupling, such as Cr$_2$O$_3$/Rh$_2$O$_3$/Cr$_2$O$_3$, pronounced electronic hybridization is observed across the entire band structure of the trilayer as seen in Fig.~\ref{fig4} in comparison to the component 2D and bilayer spectra.
Projected band structure analysis reveals extensive mixing of electronic states originating from the different 2D subsystems, such that the resulting band structure cannot be described as a simple superposition of the constituent monolayers or even the corresponding bilayer HS.
Both valence and conduction bands are affected, leading to a narrowing of the band gap by about 0.4 eV in comparison to the bilayer and the emergence of a qualitatively new electronic structure.
This behavior reflects the formation of a strongly coupled hybrid electronic system driven by robust interfacial bonding.

\begin{figure}[ht!]
    \includegraphics[width=0.5\textwidth]{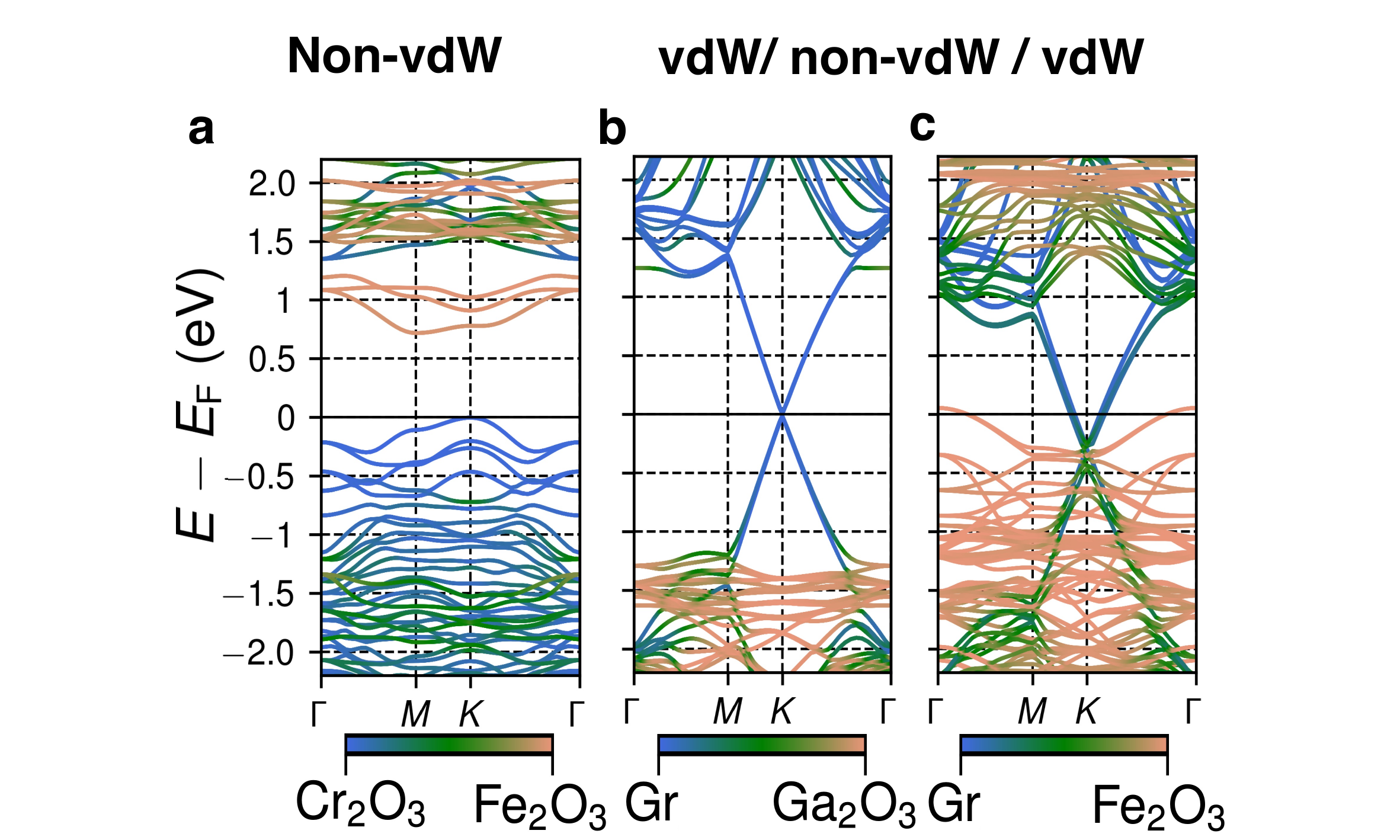}
    \caption{\small \textbf{Comparison of non-vdW and vdW/non-vdW sandwich-like band structures.}
    (\textbf{a}) Non-vdW HS (Cr$_2$O$_3$/Fe$_2$O$_3$/Cr$_2$O$_3$) band structure.
    Hybrid vdW/non-vdW/vdW HSs band structures: Gr/Ga$_2$O$_3$/Gr (\textbf{b}) and Gr/Fe$_2$O$_3$/Gr (\textbf{c}).
    }
    \label{fig5}
\end{figure}

At the same time, not all purely non-vdW sandwiches exhibit strong band hybridization over the whole spectrum.
In Cr$_2$O$_3$/Fe$_2$O$_3$/Cr$_2$O$_3$ (Fig.~\ref{fig5}(\textbf{a})), the electronic states near the band gap edges largely retain their layer-resolved character, despite the presence of strong interfacial bonding.
In this case, the primary electronic effect is an increase of the band gap by 0.3~eV relative to the bilayer HS, which can be explained by the disappearance of the surface state derived band of the encapsulated Fe$_2$O$_3$.
This highlights that strong structural coupling does not necessarily imply strong band mixing around the gap, particularly when the electronic states of the constituent layers are energetically or symmetrically mismatched.

Complementary effects are observed in hybrid vdW/non-vdW/vdW sandwiches.
For example, for Ga$_2$O$_3$ encapsulated between graphene sheets, electronic hybridization remains very weak and is largely confined to deeper valence states, while the band edges of Ga$_2$O$_3$ are preserved (Fig.~\ref{fig5}(\textbf{b})). In this regime, encapsulation primarily stabilizes the non-vdW layer and induces negligible (0.05 eV) band gap renormalization compared to the pristine 2D sheet without qualitatively altering its intrinsic electronic structure.

On the other hand, encapsulation of Fe$_2$O$_3$ by graphene leads to a more pronounced electronic response.
Charge transfer from Fe$_2$O$_3$ to graphene results in n-type doping of the graphene layers, evidenced by a downward shift of the Dirac point by approximately 0.32~eV relative to the Fermi level (Fig.~\ref{fig5}(\textbf{c})).
This doping drives graphene into a metallic state, while Fe$_2$O$_3$ becomes p-doped.
These differences demonstrate that the electronic behavior of vdW/non-vdW/vdW sandwiches is strongly dependent on the chemical and electronic properties of the encapsulated non-vdW layer.
Projected band structures for four additional trilayer HSs are provided in Fig.~S3 in the SI.

Overall, these results already demonstrate that sandwich HSs provide a versatile platform for tuning electronic properties, enabling regimes ranging from weakly perturbed, electronically decoupled layers to strongly hybridized systems with emergent electronic states.

\textbf{Magnetic Properties and Interlayer Exchange Coupling.}
The interfacial chemical bonding mediates now also a pronounced magnetic coupling across the sandwich structures.
Magnetic interactions are investigated for purely non-vdW sandwich HSs by comparing aligned, \emph{i.e.} ferromagnetic (FM) interfacial, and anti-aligned, \emph{i.e.} antiferromagnetic (AFM) interfacial configurations.
Two distinct exemplary magnetic architectures are considered.

\begin{figure}[h]
    \includegraphics[width=0.5\textwidth]{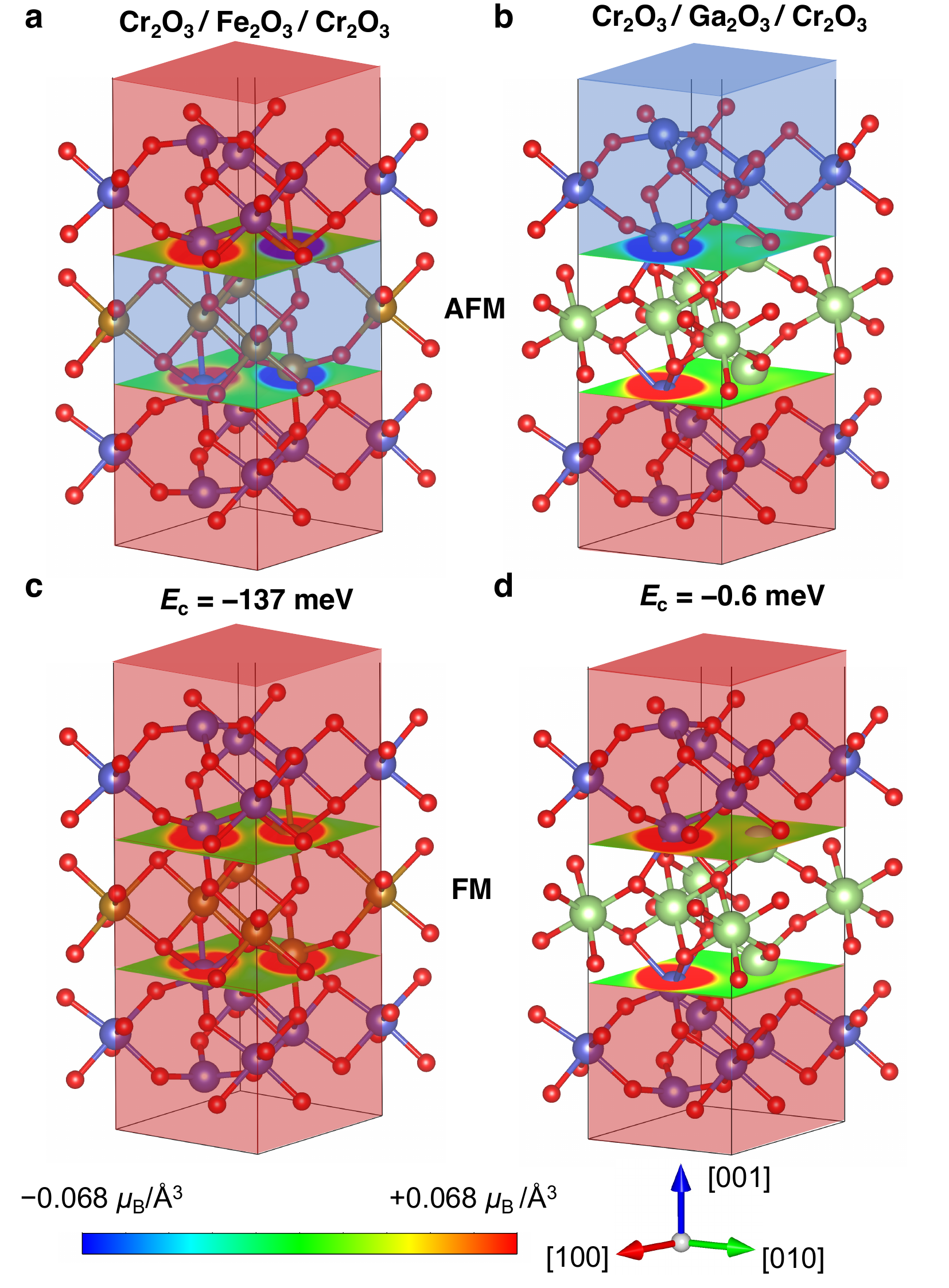}
    \caption{\small \textbf{Magnetic coupling in non-vdW sandwich-like HSs.}
    (\textbf{a}) Anti-alignment (AFM interfacial) configuration of the magnetization of the central layer with respect to the outer ones in Cr$_2$O$_3$/Fe$_2$O$_3$/Cr$_2$O$_3$.
    (\textbf{c}) Aligned (FM interfacial) configuration in the same system.
    (\textbf{b}) and (\textbf{d}) AFM and FM configurations for Cr$_2$O$_3$/Ga$_2$O$_3$/Cr$_2$O$_3$, \emph{i.e.} with a non-magnetic Ga$_2$O$_3$ spacer.
    Magnetic coupling energies $E_{\mathrm{c}}$ are also indicated.
    Positive/negative alignment of the total magnetization of a given non-vdW 2D sheet is indicated by the red/blue semitransparent boxes.
    Red and blue regions on the interfacial planes represent spin up and spin down magnetization densities, respectively.
    Atom colors as in Fig.~\ref{fig1} and Ga, light green.}
    \label{fig6}
\end{figure}

In the first case, all three constituent layers are magnetic, exemplified by the Cr$_2$O$_3$/Fe$_2$O$_3$/Cr$_2$O$_3$ sandwich.
Here, different magnetic configurations arise from alternative alignments (AFM/FM (Fig.~\ref{fig6}(\textbf{a/c}))) of the magnetization of the inner sheet with respect to the outer ones.
This leads to different alignment of the local magnetic moments across the chemically bonded interfaces, with the relative alignment within each sheet fixed.
Red and blue regions on the interfacial planes denote positive and negative spin-density contributions, revealing the spatial distribution and relative orientation of the magnetic moments on the interfacial cations.
The contrast between the two configurations reflects the change in magnetic ordering and the resulting interlayer exchange interactions.

In the second architecture, a non-magnetic Ga$_2$O$_3$ sheet separates the two magnetic Cr$_2$O$_3$ layers. In this case, the AFM and FM configurations are defined by the antiparallel and parallel alignment of the magnetic moments in the outer layers, respectively (Fig.~\ref{fig6}(\textbf{b/d})).
The magnetization density distributions show that the magnetic polarization is concentrated primarily on the Cr ions, whereas the Ga$_2$O$_3$ spacer does not carry any spin polarization.
This indicates that the spacer layer remains essentially diamagnetic while mediating the magnetic coupling between the outer layers.

Across the investigated systems, the anti-aligned configuration is energetically favored, indicating that AFM interfacial coupling is preferred.
However, the magnitude of the exchange interaction strongly depends on the type of architecture and the chemical nature of the inner layer.
To quantify this behavior, the interlayer magnetic coupling energy, \emph{i.e.} the energy difference between both configurations, $E_{\mathrm{c}}$ is evaluated.
In the fully magnetic trilayer (Cr$_2$O$_3$/Fe$_2$O$_3$/Cr$_2$O$_3$), the coupling energy is substantial reaching a value of $-$137 meV per unit cell, reflecting strong exchange mediated by the direct chemical bonding and substantial orbital hybridization across the interfaces.
Such large coupling strengths suggest robust magnetic ordering.

In contrast, when a diamagnetic spacer layer separates the magnetic sheets, the absolute value of $E_{\mathrm{c}}$ is reduced by approximately one to two orders of magnitude, typically to $-$1 to $-$10 meV per unit cell reaching $-0.6$~meV for Cr$_2$O$_3$/Ga$_2$O$_3$/Cr$_2$O$_3$ (coupling energies for Fe$_2$O$_3$/MgTiO$_3$/Fe$_2$O$_3$ and Cr$_2$O$_3$/Rh$_2$O$_3$/Cr$_2$O$_3$ systems are shown in Fig.~S4 in the SI).
This reduction signals that the spacer effectively decouples the sheets.
With such small coupling, switching of the configuration in an external magnetic field, for leveraging the arrangement for instance in a spin-polarized tunnel junction, might become feasible.

To estimate the field strength required for switching, the energy of each magnetic layer in the external field can be computed as $E=-\mathbf{m}\cdot \mathbf{B}$, where $\mathbf{m}$ is the magnetization in a given sheet and $\mathbf{B}$ is the external field.
This assumes that the magnetization of one Cr$_2$O$_3$ sheet switches as one entity and not each Cr moment individually as already successfully applied for different magnetic subunits within a single molecule-surface hybrid system~\cite{Hess_NJP_2017}.
This is motivated by the fact that the coupling between different Cr moments inside one sheet are much larger than across the spacer in the sandwich structure.
When evaluating the switching field from the energy difference of both configurations using that the magnetization of one Cr sheet is $\sim12~\mu_{\mathrm{B}}$ (with $\mu_{\mathrm{B}}$ being the Bohr magneton) one arrives at $B_{\mathrm{switch}}\sim0.43$~T.
Such moderate field values are easily achievable indicating that the investigated magnetic sandwich HSs could be used for potential functionality implementation in spintronics.

\section{Conclusions} \label{conclusions}

Our results emphasize the potential of 2D sandwich architectures in controlling the balance between property mixing and property preservation in HSs.
Purely non-vdW sandwiches show strong interfacial bonding as evidenced by the optimized structures, binding energies, and charge density difference analysis.
Absolute binding energies reach up to approximately 190~meV/\AA$^2$ per interface indicating the formation of covalent/ionic bonds accompanied by substantial geometric reconstruction and charge redistribution.
Their binding energies thus turn out to be a factor of three to eight larger than arrangements when a non-vdW sheet is placed between two vdW layers and almost an order of magnitude larger than pure vdW bonding.

In contrast, hybrid vdW/non-vdW/vdW sandwiches exhibit considerably weaker interfacial interactions, with absolute binding energies of approximately 25–30 meV/\AA$^2$ per interface.
They largely preserve the intrinsic atomic structure and electronic characteristics of the enclosed non-vdW layer, while enabling interface-driven charge transfer doping of graphene in certain cases.
This scenario thus outlines vdW encapsulation as a viable route to protect non-vdW materials from the environment to preserve their properties.

The magnetic analysis demonstrates that chemically bonded non-vdW sandwiches favour AFM interfacial alignment.
The strength of the interlayer magnetic coupling depends strongly on the sandwich composition, reaching $-137$ meV per unit cell in Cr$_2$O$_3$/Fe$_2$O$_3$/Cr$_2$O$_3$, while insertion of a diamagnetic spacer such as Ga$_2$O$_3$ between Cr$_2$O$_3$ sheets reduces the coupling to approximately $-1$ to $-10$ meV per unit cell.
These findings highlight the potential of such sandwich architectures as switchable magnetic junctions for implementing spintronic functionalities.

Overall, the ability to tune the systems between property mixing and property preservation for a range of characteristics by selecting the appropriate sandwich architecture opens new pathways for the rational design of multifunctional 2D HSs.
They might hence enable bridging the gap between strongly bonded thin films and weakly interacting vdW stacks.
Importantly, the trilayer geometry allows for stronger and more diverse electronic coupling effects than those accessible in bilayer HSs, offering additional degrees of freedom for materials and interface design.

\section{Methods} \label{methods}

HSs are generated by a custom ``AFLOW-Hetbuilder" implementation of the coincidence lattice method
following the work of Koda \emph{et al.} \cite{Koda_jpcc_2016,Kempt_hetbuilder}.
The initial structures for the non-vdW 2D sheets are sourced from the dataset of previously proposed candidates \cite{Friedrich_NanoLett_2022,Barnowsky_AdvElMats_2023}. The vdW layers are generated using the aflow slab command \cite{curtarolo:art62} on the bulk parent structures of graphene and MoS$_2$.
The components are initially placed 2~\AA\ apart in the vertical direction perpendicular to the sheets.
The sandwich-like HSs are generated via duplicating the bottom layer above the encapsulated layer from the bilayer HSs \cite{Nihei_ActaMat_2026}.

For all HSs, \DFT\ calculations including structural relaxations as well as evaluation of band structure and density of states (DOS) are performed.
The computations for the exchange-correlation functionals SCAN+rVV10 \cite{Perdew_SCAN_PRL_2015,Sabatini_rVV10_PRB_2013,Peng_PRX_2016} for relaxation and PBE+$U$ \cite{PBE,Anisimov_Mott_insulators_PRB1991,Liechtenstein1995,Dudarev1998,Grimme_JCP_2010} for electronic properties are executed with AFLOW \cite{Divilov_Aflow4_2025,Esters_CMS_2023} commanding the Vienna Ab-initio Simulation Package (\VASP) \cite{vasp_prb1996,kresse_vasp_paw} with settings according to the AFLOW standard \cite{curtarolo:art104}
and internal \VASP\ precision set to ACCURATE.
A comparison of the binding energies to results from PBE+$U$+D3 \cite{Grimme_JCP_2010} is shown in Table S2 in the SI.
Electrostatic dipole correction models for energies, potentials, and forces are included \cite{Neugebauer_PRB_1992}, along with non-spherical contributions from gradient corrections inside projector-augmented wave (PAW) spheres.

The AFLOW internal automatic determination of $k$-point sets is used in conjunction with an extension for 2D systems enabling only in-plane sampling to be released in a future version.
The number of $k$-points per reciprocal atom is set to 2500 for static calculations which is reduced to 400 for structural relaxations \cite{Nihei_ActaMat_2026}.

The charge density difference is computed as:

\begin{equation}
    n_{\mathrm{diff}}(\mathbf{r})= n_{\mathrm{HS}}(\mathbf{r}) - 2n_{\mathrm{outer}}^{\mathrm{static}}(\mathbf{r}) - n_{\mathrm{inner}}^{\mathrm{static}}(\mathbf{r}),
\end{equation}

where $n_{\mathrm{HS}}(\mathbf{r})$ is the charge density of the relaxed HS and $n_{\mathrm{outer/inner}}^{\mathrm{static}}(\mathbf{r})$ is the density of the outer/inner 2D component(s) with the same atomic positions as in the HS obtained from a single self-consistent calculation.

The binding energy $E_\mathrm{b}$ per unit area is calculated as:
\begin{equation} \label{Eb}
    E_\mathrm{b} = \frac{E_\mathrm{HS} - 2E_\mathrm{outer} - E_\mathrm{inner}}{A},
\end{equation}
where $E_\mathrm{HS}$ is the total energy of the HS, $E_\mathrm{outer}$/$E_\mathrm{inner}$ is the energy of the outer/inner 2D component(s), and $A$ is the area of the relaxed HS.
The $E_\mathrm{b}$ is therefore calculated per two interfaces in the sandwich structure and must be divided by two in order to be comparable to the binding energy of bilayer HSs \cite{Nihei_ActaMat_2026}.

All systems are rigorously checked for magnetism.
For systems formed from diamagnetic components, both spin-polarized and non spin-polarized calculations are conducted.
However, none of the systems developed a magnetic ground state significantly lower in energy than the diamagnetic one.
In non-vdW HSs, the magnetic configuration of the individual sheets is preserved and initialized as known from our previous studies \cite{Friedrich_NanoLett_2022,Barnowsky_AdvElMats_2023} of the isolated 2D components.
In addition we  try for five different sizes of all magnetic moments from 1 to 5 $\mu_{\mathrm{B}}$ \cite{Friedrich_CCE_2019}.
The resulting magnetic coupling energy $E_\mathrm{c}$ is calculated as the difference between the corresponding anitferromagnetic ($E_\mathrm{AFM}$) and ferromagnetic ($E_\mathrm{FM}$) interfacial states:
\begin{equation}
    E_\mathrm{c} = E_\mathrm{AFM} - E_\mathrm{FM}.
\end{equation}
A negative coupling energy thus indicates that the AFM alignment is energetically preferred.

We used the AI-based language model Chat-GPT to assist with the refinement of the manuscript text.

\section{Data Availability} \label{data_avail}
The primary research data of this study are available from the Rossendorf Data Repository (RODARE) via \verb|https://doi.org/10.14278/rodare.5110|~\cite{Nihei_Rodare_2026}.

\section{Author contributions} \label{contribs}

R.F. initiated and supervised the project.
A.N. built the HSs and performed all \DFT\ calculations and all post-processing.
A.N. and R.F. analyzed the data.
A.N. wrote the initial draft of the manuscript which was then refined by R.F.
Both authors contributed to the writing and discussion of the final manuscript.

\begin{acknowledgments}

The authors thank Tom Barnowsky for assisting with technical implementations and acknowledge Tom Barnowsky, Chen-Chen Er, and Balaram Thakur for fruitful discussions.
R.F. acknowledges funding from the Deutsche Forschungsgemeinschaft (DFG), project FR 4545/2-1, project ID: 513655797, and the collaborative research center ``Chemistry of Synthetic Two-Dimensional Materials'' SFB-1415, project C08, project ID 417590517 and for the “Autonomous Materials Thermodynamics” (AutoMaT) project by Technische Universität Dresden and Helmholtz-Zentrum Dresden-Rossendorf within the DRESDEN-concept alliance.
The authors thank the HZDR Computing Center, HLRS Stuttgart (HAWK cluster), and the Paderborn Center for Parallel Computing (PC2, Noctua 2 and Otus clusters) for generous grants of CPU time.

\end{acknowledgments}

\newcommand{\Ozolins}{Ozoli{\c{n}}{\v{s}}}

\clearpage
\FloatBarrier
\onecolumngrid
\begin{center}
  \Large
  \textbf{
  Supplementary Information
  }
\end{center}

\setcounter{figure}{0}
\setcounter{table}{0}
\setcounter{section}{0}
\renewcommand{\thesection}{S\Roman{section}}
\renewcommand{\thetable}{S\Roman{table}}
\renewcommand{\thefigure}{S\arabic{figure}}

\section{R\lowercase{elaxed }S\lowercase{tructures,} K\lowercase{ey} s\lowercase{tructural and} E\lowercase{nergetic} p\lowercase{roperties,} c\lowercase{omparison of} b\lowercase{and} s\lowercase{tructures, and} m\lowercase{agnetic coupling of }S\lowercase{andwich-like non-vd}W \lowercase{and }H\lowercase{ybrid vd}W/\lowercase{non-vd}W/\lowercase{vd}W HS\lowercase{s}} \label{rel_str}

    \begin{figure}[h!]
        \centering
        \includegraphics[width=0.5\columnwidth]{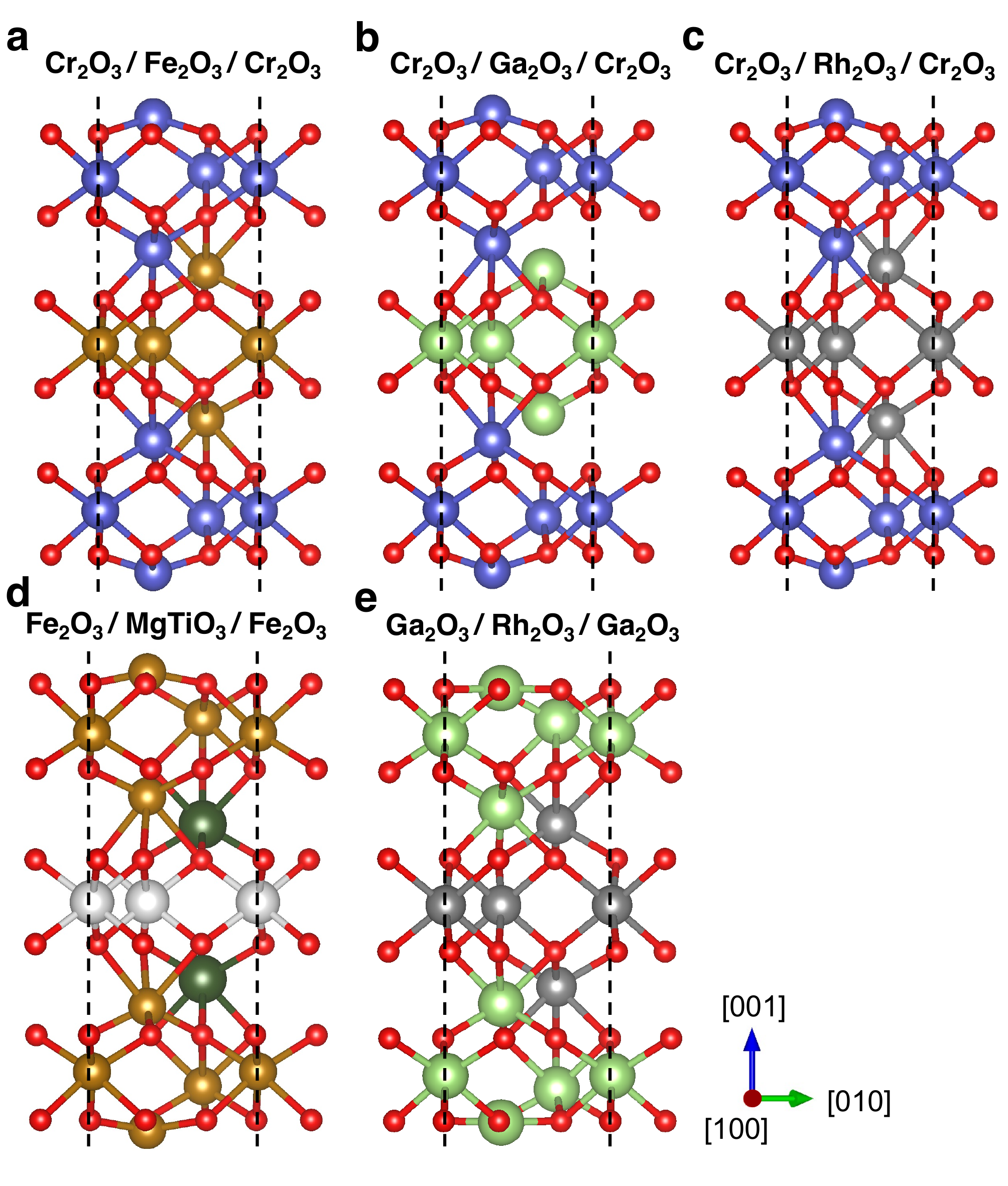}
        \caption{\small \textbf{Optimized structures of non-vdW sandwich-like HSs.} Relaxed structures of (\textbf{a}) Cr$_2$O$_3$/Fe$_2$O$_3$/Cr$_2$O$_3$, (\textbf{b}) Cr$_2$O$_3$/Ga$_2$O$_3$/Cr$_2$O$_3$, (\textbf{c}) Cr$_2$O$_3$/Rh$_2$O$_3$/Cr$_2$O$_3$, (\textbf{d}) Fe$_2$O$_3$/MgTiO$_3$/Fe$_2$O$_3$, and (\textbf{e}) Ga$_2$O$_3$/Rh$_2$O$_3$/Ga$_2$O$_3$. 
        The compass indicating crystal directions refers to all subfigures.
        } 
         \label{SI_fig1}
    \end{figure}

     \begin{figure}[h!]
        \centering
        \includegraphics[width=0.8\columnwidth]{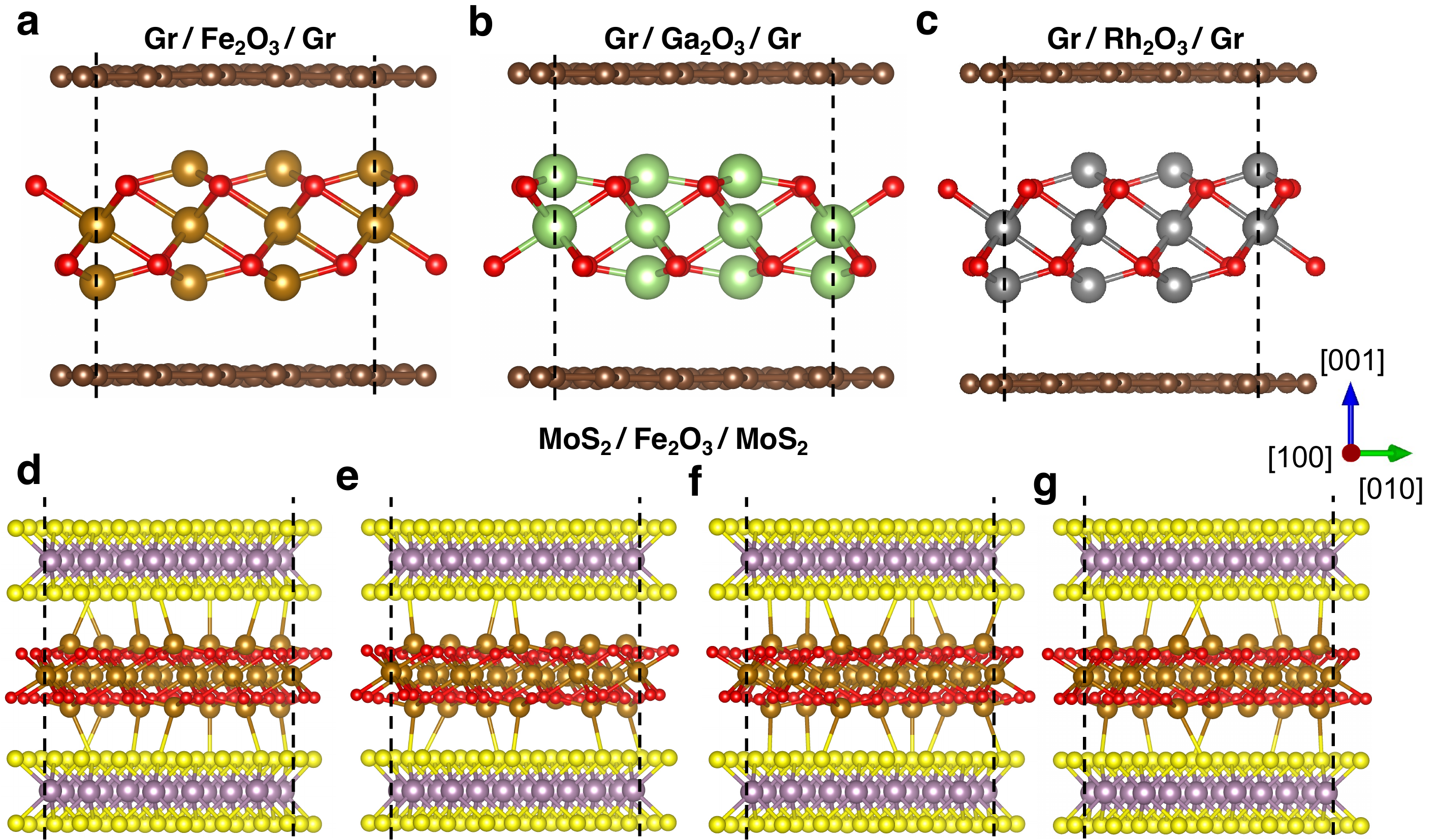}
        \caption{\small \textbf{Optimized structures of hybrid vdW/non-vdW/vdW sandwich-like HSs.} Relaxed structures of (\textbf{a}) Gr/Fe$_2$O$_3$/Gr, (\textbf{b}) Gr/Ga$_2$O$_3$/Gr, (\textbf{c}) Gr/Rh$_2$O$_3$/Gr, and MoS$_2$/Fe$_2$O$_3$/MoS$_2$ with twist angle of 4$^{\circ}$ (\textbf{d}), 17$^{\circ}$ (\textbf{e}), 42$^{\circ}$ (\textbf{f}) and 55$^{\circ}$ (\textbf{g}).  
        The compass indicating crystal directions refers to all subfigures.}
         \label{SI_fig2}
    \end{figure}

\begin{table}[h!]
\centering
\caption{Summary of key structural and energetic parameters for sandwich-like heterostructures. Numbers in brackets for MoS$_2$/Fe$_2$O$_3$/MoS$_2$ systems indicates the twist angle between individual 2D sheets. \vspace{0.1cm}}
\resizebox{\textwidth}{!}{ 
\begin{tabular}{|>{\centering\arraybackslash}m{3.5cm}|>{\centering\arraybackslash}m{3.5cm}|>{\centering\arraybackslash}m{3.5cm}|>{\centering\arraybackslash}m{3.5cm}|}
\hline
\textbf{Outer/Inner Layer} & \textbf{Interlayer Bonding Type} & \textbf{Interlayer Distance $d$ (\AA)} & \textbf{Binding Energy (meV/\AA$^2$)} \\ \hline
\hline
Cr$_2$O$_3$/Fe$_2$O$_3$ & Covalent/Ionic & 1.44--1.47 & $-379$ \\ \hline
Cr$_2$O$_3$/Ga$_2$O$_3$ & Covalent/Ionic & 1.6--1.66 & $-189$ \\ \hline
Cr$_2$O$_3$/Rh$_2$O$_3$ & Covalent/Ionic & 1.6--1.62 & $-168$ \\ \hline
Fe$_2$O$_3$/MgTiO$_3$ & Covalent/Ionic & 1.47--1.67 & $-300$ \\ \hline
Ga$_2$O$_3$/Rh$_2$O$_3$ & Covalent/Ionic & 1.37 & $-364$ \\ \hline
\hline
Gr/Fe$_2$O$_3$ & van der Waals & 2.57 & $-57$ \\ \hline
Gr/Ga$_2$O$_3$ & van der Waals & 2.82 & $-56$ \\ \hline
Gr/Rh$_2$O$_3$ & van der Waals & 2.9 & $-46$ \\ \hline
MoS$_2$/Fe$_2$O$_3$ (4$^{\circ}$) & van der Waals & 2.47 & $-58$ \\ \hline
MoS$_2$/Fe$_2$O$_3$ (17$^{\circ}$) & van der Waals & 2.48 & $-59$ \\ \hline
MoS$_2$/Fe$_2$O$_3$ (42$^{\circ}$) & van der Waals & 2.47 & $-59$ \\ \hline
MoS$_2$/Fe$_2$O$_3$ (55$^{\circ}$) & van der Waals & 2.47 & $-58$ \\ \hline
\end{tabular}
}
\label{tab:SI_extended_summary}
\end{table}

\begin{figure}[h!]
    \includegraphics[width=0.7\textwidth]{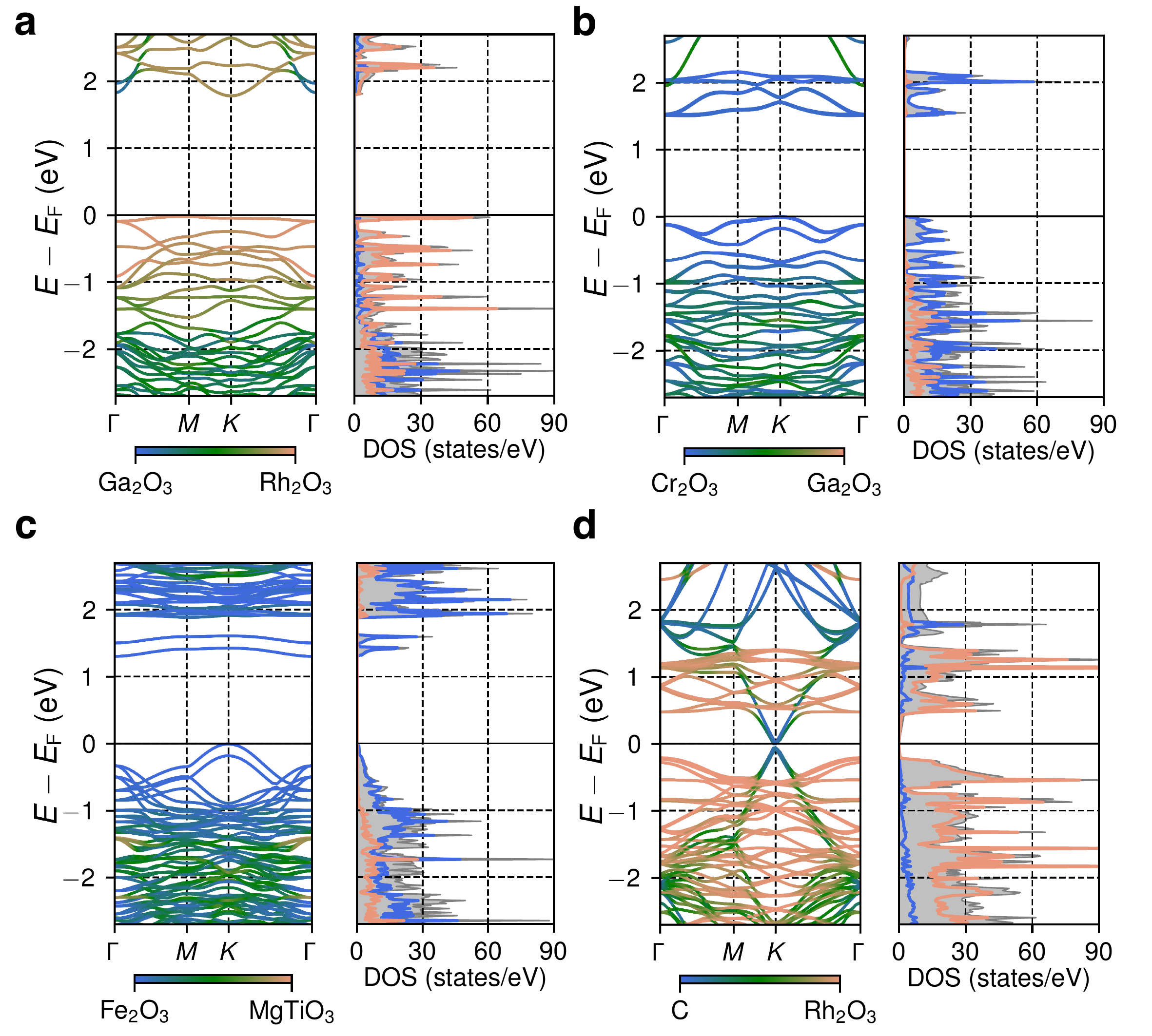}
    \caption{\small \textbf{Comparison of non-vdW and vdW/non-vdW sandwich-like band structures.} 
	(a-c) Non-vdW HSs band structures (Ga$_2$O$_3$/Rh$_2$O$_3$/Ga$_2$O$_3$, Cr$_2$O$_3$/Ga$_2$O$_3$/Cr$_2$O$_3$ and Fe$_2$O$_3$/MgTiO$_3$/Fe$_2$O$_3$). (d) Hybrid vdW/non-vdW/vdW HS (Gr/Rh$_2$O$_3$/Gr) band structure. C indicates graphene.}
    \label{SI_fig3}
\end{figure}

\begin{figure}[h!]
    \includegraphics[width=0.5\textwidth]{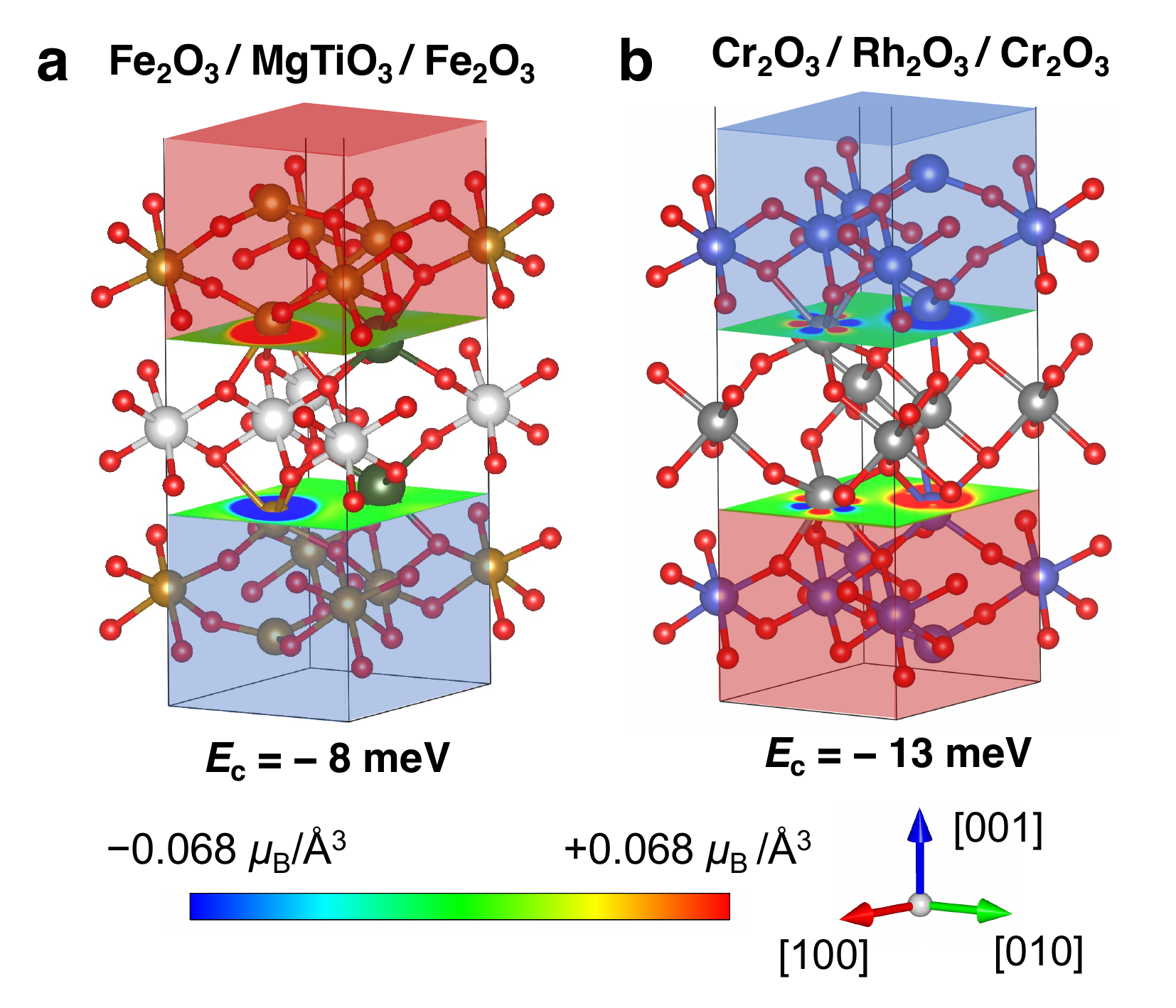}
    \caption{\small \textbf{Magnetic coupling in non-vdW sandwich-like HSs.} (a) AFM interfacial alignment of magnetic moments in Fe$_2$O$_3$/MgTiO$_3$/Fe$_2$O$_3$. 
    (b) AFM interfacial configuration for Cr$_2$O$_3$/Rh$_2$O$_3$/Cr$_2$O$_3$.
    Magnetic coupling energies $E_{\mathrm{c}}$ are also indicated.
    Red and blue regions on the interfacial planes represent spin up and spin down magnetization densities, respectively.  
    The compass indicating crystal directions refers to both subfigures.
    }
    \label{SI_fig4}
\end{figure}

\clearpage
\section{B\lowercase{enchmarking }L\lowercase{evel of }T\lowercase{heory}} \label{D3_comparison}

\begin{table}[h!]
    \caption{Comparison of interfacial binding energies $E_\mathrm{b}$ for different systems calculated with PBE+$U$ including dispersion correction (D3) \cite{Grimme_JCP_2010} as well as with SCAN+rVV10. 
    Numbers in brackets for the MoS$_2$/Fe$_2$O$_3$ systems indicate the twist angle between individual 2D sheets. }
\centering
\begin{tabular}{|c|c|c|}
\toprule
\hline
\textbf{Outer/Inner layer} & \textbf{PBE+U+D3} & \textbf{SCAN+rVV10} \\
\midrule
\hline
\hline
Cr$_2$O$_3$/Fe$_2$O$_3$ & $-343$ & $-379$ \\
\hline
Cr$_2$O$_3$/Ga$_2$O$_3$ & $-167$ & $-189$ \\
\hline
Cr$_2$O$_3$/Rh$_2$O$_3$ & $-144$ & $-168$ \\
\hline
Fe$_2$O$_3$/MgTiO$_3$ & $-238$ & $-300$ \\
\hline
Ga$_2$O$_3$/Rh$_2$O$_3$ & $-364$ & $-364$ \\
\hline
Gr/Fe$_2$O$_3$ & $-44$ & $-57$ \\
\hline
Gr/Ga$_2$O$_3$ & $-42$ & $-56$ \\
\hline
Gr/Rh$_2$O$_3$ & $-34$ & $-46$ \\
\hline
MoS$_2$/Fe$_2$O$_3$ (4$^{\circ}$) & $-66$ & $-58$ \\
\hline
MoS$_2$/Fe$_2$O$_3$ (17$^{\circ}$) & $-67$ & $-59$ \\
\hline
MoS$_2$/Fe$_2$O$_3$ (42$^{\circ}$) & $-67$ & $-59$ \\
\hline
MoS$_2$/Fe$_2$O$_3$ (55$^{\circ}$) & $-66$ & $-58$ \\
\hline
\bottomrule
\end{tabular}

\label{tab:SI_D3_comp}
\end{table}

\end{document}